# From the vapor-liquid coexistence region to the supercritical fluid: the van der Waals fluid

Hongqin Liu (刘洪勤)

Integrated High Performance Computing Branch

Shared Services Canada, Montreal, Canada

## Abstract

In this work the interface system of the van der Waals fluid is investigated by using the density gradient theory incorporated with the mean-field theory. Based on the mean-field dividing interface generated by the Maxwell construction, we propose a highly accurate density profile model for the density gradient theory, which facilitates reliable predictions of various properties for the interface region. It is found that the local intrinsic Helmholtz free energy peaks at the interface and that the maximum difference of the normal and tangential components of the pressure tensor corresponds to the maximum of the intrinsic Gibbs free energy. It is found that the entire phase space is divided into gas-like and liquid-like regions by the single line composed of the mean-field interface and the Widom line. The two-fluid feature of the supercritical fluid is hence inherited from the coexistence region. Phase diagrams extended into the coexistence region in all the temperature-pressure-volume planes are thus completed with the solutions to the vapor-liquid equilibrium problem by the van der Waals equation of state.

Emails: hongqin.liu@ssc-spc-gc.ca; hqliu2000@gmail.com

# Introduction

In the mean field theory [1] the effect of many-body interactions is approximated by an average effect and thus a many-body problem is reduced to a one-body problem. The celebrated van der Waals (vdW) equation of state (EoS) [2] is a typical example of the application of the mean-field theory in dealing with a stable vapor-liquid system. With the vdW EoS the net molecular interaction is a result of the competition between the repulsive and attractive forces. The vdW EoS set up the foundation for the vapor-liquid equilibrium (VLE) calculation as subjected to the Maxwell construction (the equal-area rule) [3]. Based on the vdW EoS, several cubic EoS's have been proposed successfully for solving VLE problems for various systems [4]. Here a VLE problem refers to calculations of the saturated pressure, vapor and liquid volumes (densities) and other related properties in the vapor-liquid existence region. With a cubic EoS and majority more complex EoS [5], there are three solutions (roots) to a VLE problem, one as the vapor volume, another, the liquid volume and third one, an "unphysical" solution. The last one has been considered as an artifact of the mean-field theory and thus discarded.

In the coexistence (interfacial) region, or the interface area, the unstable system is composed of heterogeneous nanoscale clusters and the equilibrium thermodynamic (the mean-field theory alone) fails to apply. By using a perturbation approach van der Waals proposed the density gradient theory [1], which is later enriched and finalized by Cahn and Hilliard [6]. In the density gradient theory, a position-dependent density function, known as the density profile, is introduced to bridge the discontinuous bulk densities so that a continuous density variable can be used to define any state in the region. Various local (position-dependent) properties are composed of two parts: (1) a homogeneous contribution from the mean-field theory with the local density; (2) a heterogeneous counterpart, which is expressed in terms of the density gradients (derivatives). The basic assumption is that conventional (classic) thermodynamics formalisms hold in the interface area as the heterogeneous contribution is considered.

The density profile plays a key role here and it can be obtained from theories [7-10] or from computer simulations [11-15]. Since a position variable (for a planar interface) is introduced, the origin of this variable, known as the dividing interface, needs to be defined. The definition of the dividing interface is arbitrary from a macroscopic point of view. Traditionally, the algebraic mean density (average of the bulk densities) is defined as the Gibbs dividing interface [1]. There is a basic shortcoming with this definition: the decay length of the vapor side (boundary) is the same as that of the liquid phase, which is obviously against physical expectation [9]. Another issue is that the positional dependence of the derivatives of the density profile based on the classic dividing interface is incorrect and leads to incorrect predictions of the difference between the normal and tangential components of the pressure tensor reported by computer simulations [15].

As temperature rises the two-phase coexistence region diminishes and finally vanishes at the critical point. A system above the critical point is known as the supercritical fluid. Traditionally, the stable supercritical fluid is considered to be a uniform phase. However, in the last decades or so, some outstanding features have been discovered in the supercritical region. In particular, a line defined with the locus of the local-maximum heat capacity, $C_p$, known as the Widom line [16], is found to divide a supercritical area into gas-like and liquid-like regions [17-21]. Another characteristic line that crosses the coexistence curve and extends to a supercritical region is the Fisher–Widom line [22]. This line crosses over the coexistence curve at a spot below the critical point. Besides the Fisher-Widom line, the relationship between the behaviors in the supercritical region and those in the interface area is rarely addressed [23]. Moreover, the Widom line is currently considered to be a smooth continuation of the saturated pressure in the pressure-temperature space [23], while in the pressure-volume (density) and the temperature-density spaces, the extension of the Widom line becomes bifurcated.

In this work, we revisit the density gradient theory and the mean-field theory for the vapor-liquid interface system of the van der Waals fluid. Firstly, we define a dividing interface generated by the mean-field vdW EoS, then a new density profile model will be proposed based on a more accurate solution to the governing partial differential equation for the interface region [9]. By using the new density profile expression, various intrinsic properties will be obtained and discussed for exploring the interface system. Finally, the relationship between the properties of supercritical fluid and the coexistence region will be addressed.

## On the roots of the van der Waals equation of state

We start with the VLE problem with the vdW EoS [2] as subjected with the Maxwell construction [3]. Although this is an outdated subject, the approach proposed in this work and results obtained are novel and useful for other cubic EoS and hence provided here. The main goal is to obtain a new Gibbs dividing interface for succeeding applications. The vdW EoS can be written in a reduced form:

$$P_r = \frac{8T_r}{3v_r - 1} - \frac{3}{v_r^2} \quad (1)$$

where the reduced pressure is defined as $P_r = P/P_c$, the temperature, $T_r = T/T_c$ and the molar volume $v_r = v/v_c$. In the reduced form, the two constants, $a$ and $b$, appeared in the attractive and repulsive terms, respectively, are related to the critical constants, $T_c$ and $P_c$. By applying the pressure equilibrium condition, $P(T, v_{rL}) = P(T, v_{rv})$, we obtain:

$$[8T_r v_{rv}^2 - 3(3v_{rv} - 1)]v_{rL}^2 - (3v_{rv} - 1)^2 v_{rL} + v_{rv}(3v_{rv} - 1) = 0 \quad (2)$$

In above equations, the subscript "L" and "v" refer to the liquid and vapor phases, respectively. For obtaining saturated volumes and pressure, the chemical potential equilibrium condition, $\mu_G(T, v_{rL}) = \mu_L(T, v_{rv}) = \mu_r^e(T)$, has to be imposed. Equivalently, one can use the Maxwell construction [3]:

$$P_r^e = \frac{1}{v_{rv} - v_{rL}} \int_{v_{rL}}^{v_{rv}} P_r \, dv_r \quad (3)$$

More details are presented in Appendix A, Figure A1 and Figure A2. From Eq.(2), we see that the liquid ($v_{rL}$) and the third solution ($v_{rM}$), are related to the vapor volume ($v_{rv}$) by the following equation:

$$v_{rM|L} = \frac{(3v_{rv} - 1)^2 \pm Q_v}{16T_r v_{rv}^2 - 6(3v_{rv} - 1)} \quad (4)$$

where the notation in the subscript "$M|L$" refers to $v_{rM}$ and $v_{rL}$, corresponding to "$\pm$" on the right hand side of the equation, respectively, and

$$Q_v = (9v_{rv}^2 - 1)\left[1 - \frac{32T_r v_{rv}^3}{(3v_{rv} + 1)(9v_{rv}^2 - 1)}\right]^{\frac{1}{2}} \quad (5)$$

From the Maxwell construction, Eq.(3) and EoS, Eq.(1), we obtain:

$$P_r^e = \frac{8T_r}{3(v_{rv} - v_{rL})} \ln\left(\frac{3v_{rv} - 1}{3v_{rL} - 1}\right) - \frac{3}{v_{rL} v_{rv}} \quad (6)$$

From Eq.(1) and Eq.(6) after some rearrangements, we have [24]:

$$\ln \frac{3v_{rv} - 1}{3v_{rL} - 1} = \frac{v_{rv} - v_{rL}}{v_{rG} + v_{rL}}\left(\frac{3v_{rv}}{3v_{rv} - 1} + \frac{3v_{rL}}{3v_{rL} - 1}\right) \quad (7)$$

From Eq.(2), Eq.(4) and Eq.(7), all three volumes can be obtained. By replacing $v_{rL}$ in Eq.(7) with Eq.(4), we see that the VLE calculation with the vdW EoS is equivalent to solving one-unknown ($v_{rv}$) equation, which can be easily achieved with a nonlinear equation solver, such as the Excel Solver. This approach can be applied to other cubic EoS, namely solving a VLE problem with a cubic EoS is reduced to solving one-unknown non-linear equation [25]. At the same time, we have the solution for $v_{rM}$ from Eq.(4) at the given temperature. Notable features of $v_{rM}$ are: $P_r^e = P_r(v_{rL}) = P_r(v_{rM}) = P_r(v_{rv})$ and $\mu_L(T, v_{rM}) \neq \mu_r^e(T)$, and the last one tells that the system at $v_{rM}$ is not in equilibrium or unstable state. The trajectory of $v_{rM}$ in the coexistence region is called the Maxwell crossover, or the M-line.

For the vdW EoS, some simple relations can be obtained between the saturated volumes and pressure. By rewriting Eq.(2) with $v_{rv}$ being replaced by $v_{rM}$, we have the following solutions:

$$v_{rv|L} = \frac{(3v_{rM} - 1)^2 \pm Q_M}{16T_r v_{rM}^2 - 6(3v_{rM} - 1)} \quad (8)$$

where $Q_M$ is defined the same way as Eq.(5) by replacing $v_{rv}$ with $v_{rM}$. With some simple algebra, from Eq.(4) and (8) we have

$$\rho_{rL} + \rho_{rM} + \rho_{rv} = 3 \quad (9)$$

where the reduced density, $\rho_r = 1/v_r$. Meanwhile the saturated volumes have a simple relation with temperature and with the equilibrium pressure, respectively:

$$8T_r = (3 - \rho_{rM})(3 - \rho_{rL})(3 - \rho_{rv}) \quad (10)$$

$$\rho_{rL}\rho_{rv}\rho_{rM} = P_r^e \quad (11)$$

Eq.(9), Eq.(10) and Eq.(11) are remarkable: they provide the simple relations between saturated densities, equilibrium pressure and temperature. In addition, Eq.(8) and Eq.(9) tell that if we know the M-line or the diameter of the coexistence curve then the saturated densities and hence pressure (Eq.(6)) are all known. The same approach can be extended into any cubic EoS [25].

Some other useful relations can also be derived. For instance, the diameter of the coexistence curve is defined as [26]: $d_\sigma = (\rho_{rL} + \rho_{rv})/2$, therefore, $\rho_{rM} = 3 - 2d_\sigma$. Using the diameter as a tool to study the coexistence curve (hence VLE) has been a long time effort. In general, the diameter is related to the critical exponents [26], $d_\sigma = 1 + D_\beta|\tau_0|^{2\beta} + D_\alpha|\tau_0|^{1-\alpha} + D_1\tau_0 + \cdots$, where $\tau_0 = (T - T_C)/T_C$, $\alpha$ and $\beta$ being the critical exponents. From Eq.(9), we have:

$$\rho_{rM} = 1 - 2D_\beta|\tau_0|^{2\beta} - 2D_\alpha|\tau_0|^{1-\alpha} - 2D_1\tau_0 + \cdots \quad (12)$$

For the vdW EoS, $\alpha = 0$, $\beta = 1/2$ 21. Eq.(12) tells that the information of the second-order phase transition is embedded in the M-line. Therefore, with a cubic EoS, two solutions ($\rho_{rL}, \rho_{rv}$) are related to the first-order transition, and the third one is related to the second-order transition. By the way, the mean-field order parameter [7] can be expressed in terms of $\rho_{rM}$: $\varphi = \rho_{rL} - \rho_{rG} = \sqrt{(3 + \rho_{rM})^2 - 32T_r(3 - \rho_{rM})^{-1}}$.

The above procedure provides exact solutions numerically. For obtaining an analytical expression (approximation), we take advantages of Eq.(9). The details are presented in Appendix A and the result reads:

$$\rho_{rM} = 1 - \frac{4t_0}{5} - \frac{256}{875}t_0^2 - \frac{272}{3125}t_0^3 + 0.06976t_0^5 + 0.20804t_0^7 \quad (13)$$

where $t_0 = 1 - T_r$. Eq.(13) provides exact leading three derivatives and has an total absolute average deviation of only 0.003% in the temperature range: $1.0 \geq T_r \geq 0.3$ as compared with the exact solutions. From Eq.(13), analytical expressions for $\rho_{rL}$ and $\rho_{rv}$ can be obtained from Eq.(8), and the saturated pressure from Eq.(6). By the way, at low temperature range, $T_r < 0.3$, analytical expressions are also available (see Appendix A).

## Theoretical background for the vapor-liquid interface system

In this section, we present some theoretical background and major results at macroscopic (phenomenological) and microscopic (position-dependent) levels for succeeding calculations and analysis. Throughout this work, a planner interface perpendicular to the surface with one coordinate, $z$, is considered. We start with the grand potential ($\Omega$) for an interface system [1,7]:

$$\Omega = -PV + \gamma \mathcal{A} \quad (14)$$

where $P$ is the pressure, $V$, the total volume (the lower case, $v = 1/\rho$, refers to molar volume), $\mathcal{A}$, the interfacial area (normal to $z$) and $\gamma$, the surface tension.

The above definition is independent of the position of the Gibbs dividing interface defined at $z = z_0$ [1,7]. The surface tension can be calculated from the position-dependent pressure difference [1]:

$$\gamma = \int_{-\infty}^{\infty} \left(\frac{d\gamma}{dz}\right) dz = \int_{-\infty}^{\infty} [p_N(z) - p_T(z)] dz \quad (15)$$

where $p_N(z)$ and $p_T(z)$ are the normal (to the interface) and the tangential components of the pressure tensor, respectively. While Eq.(14) provides the relationships between surface tension and other macroscopic properties, Eq.(15) presents a relation between the local property, $p_N(z) - p_T(z)$, and surface tension. The adsorption, $\Gamma$, on the interface can be calculated from the density profile, $\rho(z)$. For the vapor and liquid phase, respectively [1,7]:

$$\Gamma_v = \int_{-\infty}^{z_0} [\rho(z) - \rho_v] dz,$$

$$\Gamma_L = \int_{z_0}^{\infty} [\rho(z) - \rho_L] dz \quad (16)$$

where the subscripts, $v$ and $L$ refer to the vapor and liquid phases, respectively. $\rho_v$ and $\rho_L$ are the saturated (equilibrium) bulk densities. Apparently, $\Gamma_v > 0$, namely the adsorption at the vapor side is an excess, and $\Gamma_L < 0$, the adsorption at the liquid side is a deficit. In the conventional (classic) model, the Gibbs dividing surface is defined such that

$$\Gamma_v + \Gamma_L = 0 \quad (17)$$

and the algebraic mean density is adopted at $z = z_0$:

$$\rho(0) = \frac{\rho_v + \rho_L}{2} \quad (18)$$

Before moving on, we need to make a few remarks on Eq.(17) and Eq.(18). Macroscopically, the excess free energy of an interface is given by [1]: $F^s = \gamma \mathcal{A} + \mu n^s$ where $\mu n^s$ represents the sum of chemical potential multiplied by the excess surface density. If the surface tension is defined as $\gamma = F^s/\mathcal{A}$, then the dividing interface should be chosen in such a way that $\mu n^s = 0$ and this is where the equal-molar definition, Eq.(18), comes into play [1]. For this reason, historically Eq.(18) has been adopted for various density profile models [1,27]. However, if the grand potential, Eq.(14), is used, such a restriction is not required [1,7]. Moreover, with the density gradient theory, as the free energy is expressed as Eq.(19) below, the restriction, Eq.(17), (hence Eq.(18)) is not required either [1]. By the way, it can be proved that the total adsorption $\sum_i \Gamma_i d\mu_i$ (multicomponent) is invariant with the position change of the dividing interface [1]. In summary, the choice of the interface is arbitrary in the context of the density gradient theory [1,7], which makes a different choice physically acceptable.

In the following we briefly summarize some major equations for the Helmholtz free energy (the total free energy), $\mathcal{F}$, the local Gibbs free energy, $\mathbf{G}[\rho(z)]$ and related properties for the vapor-liquid interface system. Some details are included to facilitate the succeeding analysis and discussions. Full coverages can be found from Refs.[1,6,7,10].

According to the density gradient theory [6], for a pure system the total free energy can be obtained from the free energy density $f(\rho)$ by integration over the entire volume of the system, $V$:

$$\mathcal{F} = V \int f(\rho) \, dV = V \int [f_0(\rho) + k_1 \nabla^2 \rho + k_2 (\nabla \rho)^2 + \cdots] \, dv \quad (19)$$

where the coefficients, $k_1 = [\partial f/\partial \nabla^2 \rho]_0$, $k_2 = [\partial^2 f/(\partial \langle \nabla \rho \rangle)^2]_0$, are functions of the uniform density. In Eq.(19) the local (position-dependent) free energy density, $f(\rho) = a(\rho)\rho$, is expanded about the free energy density with the uniform density, $f_0(\rho) = a_0(\rho)\rho$, where $a_0$ represents the Helmholtz free energy of the uniform fluid. All odd-order terms vanish since $f(\rho)$ is a scalar and it must be invariant with respect to the direction of the gradient [6,7]. By applying the divergence theorem to the second derivative term and choosing a boundary such that $\nabla \rho \cdot \boldsymbol{n} = 0$ (where $\boldsymbol{n}$ is the unit vector) we have the well-known total free energy expression in the framework of the density gradient theory:

$$\mathcal{F} = \mathcal{A} \int_{-\infty}^{\infty} \left[ f_0(\rho) + \frac{1}{2} m \left( \frac{d\rho}{dz} \right)^2 \right] dz \quad (20)$$

where we only consider a flat interface with one direction, $z$ and $\nabla = d/dz$ etc.

From the above treatment, we see that Eq.(20) is effectively exact up to the 3rd order by omitting the 4th and higher-order terms. This important fact inspires us to revisit the density profile model discussed in the next section.

In Eq.(20), the coefficient, $m$, is related to $k_1$ and $k_2$ as shown by Eq.(19) and known as the influence parameter [1,6,7]. The influence parameter can be evaluated from the direct correlation function by the following relationship [7,9]:

$$m(\rho) = \frac{k_B T}{6} \int_0^{\infty} c(r,\rho) r^2 d\vec{r} \quad (21)$$

where $k_B$ is the Boltzmann constant and $c(r,\rho)$, the direct correlation function. For most applications, the density on the left hand side of Eq.(21) is treated as position-independent for a stable fluid. For a pure and stable fluid under normal pressure the density is uniquely dependent on temperature, $\rho(T)$, and therefore the influence parameter can also be considered as temperature-dependent, $m(T)$. In some cases, this parameter is even roughly treated as a constant [1,7].

For an interface system to reach equilibrium, the total free energy, Eq.(20), is minimized, and one gets:

$$\frac{m}{2} \frac{d}{dz} \left[ \left( \frac{d\rho}{dz} \right)^2 \right] = \frac{d\Omega}{dz} \quad (22)$$

Upon integration, by noticing $\frac{d\rho}{dz} \to 0$ and $\Omega \to \Omega_b$ as $z \to \infty$, where the subscript "b" refers to the bulk fluid:

$$\frac{m}{2} \left( \frac{d\rho}{dz} \right)^2 = \Delta\Omega = \Omega - \Omega_b = \Omega + P^s \quad (23)$$

where, $\mu(\rho) = df_0/d\rho$ and at equilibrium:

$$\Omega = [a_0(\rho(z)) - \mu^s]\rho(z) \quad (24)$$

$P^s$ and $\mu^s$ are the pressure and the chemical potential at saturated (equilibrium) condition, respectively. The free energy $a_0(\rho(z))$ can be evaluated by a mean-field EoS with $\rho(z)$ as the density. Then the surface tension can be calculated by the following [1]:

$$\gamma = \sqrt{2m} \int_{\rho_v}^{\rho_L} \sqrt{\Delta\Omega} \, d\rho \quad (25)$$

From Eq.(23), we also have

$$\gamma = m \int_{-\infty}^{\infty} \left( \frac{d\rho}{dz} \right)^2 dz = 2 \int_{-\infty}^{\infty} \Delta\Omega(\rho) \, dz \quad (26)$$

Since the direct correlation functions in Eq.(21) are difficult to obtain for realistic fluid, Eq.(25) and Eq.(26) can be used to evaluate the influence parameter from surface tension data. In most cases Eq.(26) is adopted to estimate the influence parameter.

Eq.(25) or Eq.(26) tells that the dividing interface will not affect the macroscopic surface tension. With the classic model, Eq.(18), the prediction of the surface may be acceptable. However, if we consider Eq.(15) and Eq.(26), the position-dependent pressure difference, $p_N(z) - p_T(z)$, depends on the local property, $\left(\frac{d\rho}{dz}\right)^2$ and accuracy of the derivative, $\frac{d\rho}{dz}$, is also crucial. This is why we need a new density profile model.

For our succeeding analysis, we present an important equation for the Gibbs free energy. The generic grand potential $\Omega_n$ and the free energy $\mathcal{F}_n$ have the following relation [7]:

$$\Omega_n[n(z)] = \mathcal{F}_n[n(z)] - N\mathbf{G}[\rho(z)] \quad (27)$$

where $N$ is the total number of particles and $\mathbf{G}$ is the Gibbs free energy. For an interface system, both $\Omega_n$ and $\mathcal{F}_n$ are functional of the generic density profile, $n(z)$. As the system reaches to equilibrium, $n(z) \to \rho(z)$, $\Omega_n[n(z)] \to \Omega[\rho(r)]$, $\mathcal{F}_n[n(z)] \to \mathcal{F}[\rho(z)]$, and $\Omega_n[n(z)]$ reaches a minimum value, therefore Eq.(27) yields:

$$\left.\frac{\delta\Omega_n[n(z)]}{\delta n(z)}\right|_{\rho(z)} = \left.\frac{\delta\mathcal{F}_n[n(z)]}{\delta n(z)}\right|_{\rho(z)} - \mathbf{G}[\rho(z)] \quad (28)$$

At equilibrium, the derivative of the generic free energy in Eq.(28) is the same as that of $\mathcal{F}(\rho(z))$ with respect to $\rho(z)$. Throughout this work, after Ref.[7,10], we define an intrinsic (local) property of an interface system as the sum of a mean-field homogeneous contribution and a heterogeneous counterpart (the density gradients), such as the intrinsic free energy given by the integrand of Eq.(20). From Eq.(20) and (28), we finally have the intrinsic Gibbs free energy [10]:

$$\mathbf{G}[\rho(z)] = \mu(\rho(z)) - m\frac{d^2\rho}{dz^2} - \frac{1}{2}m'\left(\frac{d\rho}{dz}\right)^2 \quad (29)$$

where $m' = dm/d\rho$. The homogeneous contribution, $\mu(\rho(z))$, can be calculated the same way as the free energy, $a_0(\rho(z))$, by using a mean-field EoS. Obviously, applications of Eq.(29) and the intrinsic free energy (the integrand of Eq.(20)) rely on an accurate expression for the density profile, $\rho(z)$, and numerical results from computer simulations require extra avenue for the derivatives.

Eq.(29) is an important result for exploring the interfacial region, but it is seldom addressed in the literature, unfortunately. Yong et al.[10] discussed it with some hypothetical scenarios without the involvement of the density profile and any specific expression for $\mu(\rho(z))$. In this work, this equation will be employed for the vdW fluid.

For applications of the density gradient theory, Eq.(20) and Eq.(29), etc., the density profile, $\rho(z)$, and its derivatives, $d\rho/dz$ and $d^2\rho/dz^2$ are required and an analytical function can best serve the purposes. Several analytical expressions have been proposed [1,27] with the Gibbs dividing interface, Eq.(18). For a density profile model, the following boundary conditions should be imposed [1,7]:

$$\left.\begin{array}{l} z - z_0 = -\infty, \ \rho(z) = \rho_v \\ z - z_0 = \infty, \ \rho(z) = \rho_L \\ \left.\dfrac{d^n\rho(z)}{dz^n}\right|_{z\to\pm\infty} = 0 \end{array}\right\} \quad (30)$$

A simple "classic" model that meets all the above conditions reads [1,7,15,27]:

$$\rho(z) = \frac{\rho_v + \rho_L}{2} + \frac{\rho_L - \rho_v}{2}\tanh\left(\frac{z - z_0}{D}\right) \quad (31)$$

where the parameter $D$ is related to the thickness of the interfacial region for a planar surface. Obviously, the algebraic-mean interface defined by Eq.(18) is adopted in Eq.(31). It is straightforward to prove that Eq.(31) satisfies the Gibbs dividing interface definition, Eq.(17), which demands that the "thickness" of the vapor layer equals to that of the liquid layer, $\int_{z_0}^{-\infty} dz = \int_{z_0}^{\infty} dz$.

A fundamental issue is that the equal "thicknesses" (decay length) at the vapor and the liquid sides given by Eq.(18) apparently contracts with the physical expectations since the decay toward the gas bulk phase is more rapid than to the liquid bulk phase [7,15]. Consequently, all the position-dependent properties based on Eq.(31) suffers a basic drawback. In this work, based on a novel solution to the governing differential equation for the density profile [9], a different model is proposed (see Appendix B):

$$\rho(z) = A + \frac{B\tanh\left(\dfrac{z - z_0}{D_v}\right)}{1 + C\tanh\left(\dfrac{z - z_0}{D_L}\right)} \quad (32)$$

where the coefficients $A, B$ and $C$ are dependent on the bulk fluid densities, $\rho_v$ and $\rho_L$. $z_0$, $D_v$ and $D_L$ are the parameters of the model to be determined from computer simulation data. It is straightforward to prove that if the conditions given by Eq.(18) and Eq.(30) are imposed to Eq.(32) the classic model, Eq.(31) is recovered, where $C = 0$ is a result of imposing the condition Eq.(18).

An apparent feature of the new model is that two parameters, $D_v$ and $D_L$, are introduced to separate the vapor side from the liquid side within the interface region. This is introduced empirically, and therefore Eq.(32) is semi-empirical. In addition, a new dividing interface is required to incorporate with the separation. Here we define the Maxwell crossover, M-line, Eq.(13), as the mean-field, namely:

$$z - z_0 = 0, \ \rho(z) = \rho_M \qquad (33)$$

where the expression for $\rho_M$ is given by Eq.(13). Combining Eq.(32) and Eq.(33) with Eq.(30) yields the new density profile:

$$\rho(z) = \rho_M + \frac{2(\rho_L - \rho_M)(\rho_M - \rho_v)\tanh\left(\frac{z-z_0}{D_v}\right)}{\rho_L - \rho_v + (2\rho_M - \rho_v - \rho_L)\tanh\left(\frac{z-z_0}{D_L}\right)} \qquad (34)$$

The parameters, $D_v$, $D_L$ and $z_0$ can be obtained by fitting the model with computer simulation data. Compared with the classic model, Eq.(31), the new model separates the vapor side of the interface from the liquid side and hence it overcomes the shortcoming of the classic model and is physically favorable. From the analytical expression, we can easily obtain the derivatives, $d^n\rho(z)/dz^n$. The first two derivatives to be used in this work are provided in Appendix B.

**Relations between the dividing interface and the Widom line of the vdW fluid**

Up to this point, all the notations follow conventional ones for generic descriptions. In the following all the calculations are specific for the vdW fluid and we use dimensionless quantities indicated with the superscript "*". For example, $\rho^* = N\sigma^3/V$, $T^* = T/k_B\epsilon$, $P^* = P\sigma^3/k_B\epsilon$, $a^* = F/\epsilon$, $\Gamma^* = \Gamma\sigma^2$, $\gamma^* = \gamma\sigma^2/\epsilon$ etc., where $k_B$ is the Boltzmann constant, $\sigma$, the particle diameter, $\epsilon$, the energy parameter. The critical point is defined with the values listed in Table A1 (Appendix A). Hence the reduced properties are defined as $\rho_r = \rho^*/\rho_c^*$, $T_r = T^*/T_c^*$, $P_r = P^*/P_c^*$.

First of all, we need to determine the parameters in Eq.(34). For the vdW fluid, reported data for the density profile are scare [28]. For this research, new data were provided by Dr. Mejia (2021) [29] by using the approach reported in Ref. [11], which include density profile and the pressure difference, $p_N(z) - p_T(z)$, in the temperature range, $T^* = 0.15 \sim 0.25$ ($T_r = 0.5 \sim 0.84$). From the density profile data, the parameters of Eq.(34) are correlated as functions of temperature:

$$D_v = 6.4941T^{*2} - 5.05973T^* + 1.86935 \qquad (35a)$$

$$D_L = 7.33069T^{*2} - 5.59661T^* + 2.27538 \qquad (35b)$$

$$z_0 = 108.215T^{*2} - 175.413T^{*2} + 107.679T^* - 20.9173 \qquad (35c)$$

The same data are also used to fit the classic model, Eq.(31), and we get:

$$D = 26.1467T^{*3} - 43.2617T^{*2} + 26.3455T^* - 4.3236 \qquad (36a)$$

$$z_0 = 118.477T^{*2} - 196.315T^{*2} + 121.352T^* - 23.0543 \qquad (36b)$$

The surface tension data is also provided by Mejía [29] and is fitted with an empirical equation

$$\gamma^* = 0.257(1 - T_r)^{1.466} \qquad (37)$$

The influence parameter is then obtained from Eq.(26) and fitted with a simple linear function:

$$m = -1.13948T^* + 2.2788 \qquad (38)$$

For applying Eq.(29), we need to calculate the slope $dm/d\rho$, where, as discussed, the density cannot be $\rho(z)$. In the interface region, at a given temperature, there are two saturated (bulk) densities and the derivatives, $dm/d\rho_v$ and $dm/d\rho_L$, have opposite signs. Choosing any of them is not convincing. Considering the fact that the M-line, Eq.(13), is uniquely dependent on temperature (where $T_r = T^*/T_c^*$) we use $dm/d\rho_M$ in place of $dm/d\rho$. Hence the following correlation is obtained:

$$m = -1.7\rho_M^* + 2.22 \qquad (39)$$

which gives $dm/d\rho = dm/d\rho_M = -1.7$. The excess free energy of the vdW EoS reads [30]:

$$\tilde{f}^{res} = -\ln\rho^* - \ln(\rho^{*-1} - b) - a\tau\rho^* \qquad (40)$$

where $\tilde{f}^{res} = f^{*res}/T^*$, $\tau = 1/T^*$, $a$ and $b$ are the vdW constant and they are related to the critical constant: $b = 1/(3\rho_C^*)$, $a = 27P_c^* b^2$. The chemical potential is given by

$$\mu^* = \partial(\rho^* \tilde{f})/\partial \rho^* = T^*(ln\rho^* + \tilde{f}^{res} + 1 + \partial \tilde{f}^{res}/\partial \rho^*) \quad (41)$$

With the vdW EoS, we have an analytical expression for the grand potential:

$$\Delta\Omega = \frac{1}{2}[p_N(z) - p_T(z)] = [f_0(\rho^*(z)) - \mu^{*s}]\rho^*(z) + P^{*s} \quad (42)$$

where the saturated pressure, $P^{*s}$. The equilibrium chemical potential, $\mu^{*s}$, can be obtained from the $P^* \sim \mu^*$ plane (Figure A2). The homogeneous contribution in Eq.(42) reads:

$$f_0^*(\rho^*(z)) = -T^*\left\{1 + ln\left[T^{*\frac{3}{2}}(\rho^{*-1}(z) - b)\right]\right\} - a\rho^*(z) \quad (43)$$

The residual free energy $f^{*res} = f_0^*(\rho^*(z)) - f_0^{*id}$ where $f_0^{*id} = -T^*\{1 + ln[T^{*3/2}(\rho^{*-1})]\}$. The intrinsic Gibbs free energy can be calculated with the following:

$$\mathbf{G}^*[\rho(z)] \approx T^* ln\rho^* + T^* \tilde{f}^{res} + \frac{P^*}{\rho^*} - m\frac{d^2\rho^*}{dz^2} \quad (44)$$

where the last term with $m'$ in Eq.(29) turned out to be negligible. Eq.(44) is an important relation since it allows to quantitatively calculate the Gibbs free energy for an interface system. Yang et al. [10] addressed it by using some hypothetic scenarios without involving any numerical calculations of free energy and density profile. Otherwise, it is rarely discussed in the literature. The intrinsic free energy is given by Eq.(20) and reads:

$$\mathbf{f}[\rho(z)] = f_0^*(\rho^*(z)) + \frac{1}{2}\frac{m}{\rho^*(z)}\left(\frac{d\rho^*}{dz}\right)^2$$
$$\equiv f_0^*(\rho^*(z)) + \mathbf{f_2} \quad (45)$$

By recalling that all the above equations, Eq.(44) and Eq.(45), are exact up to the 3rd order while omitting the 4th and higher order terms, the results presented below are accurate enough for our analysis.

Now we turn to the supercritical region. Numerous studies on various supercritical fluids can be found in the literature [17-23]. One of the most important findings is that the Widom line [16] divides a supercritical area into liquid-like and gas-like regions. Some other characteristic lines are also found [20,31]. The original definition of the Widom line is the local maxima of correlation length [16]. Since the later cannot be measured directly some response functions are used instead and there are multiple possibilities [31]. Here we are only interested in the Widom line defined as the locus of local maximum of the isobaric heat capacity in a constant pressure process, namely [21]:

$$(\partial C_p/\partial T)_p = 0 \quad (46)$$

Fortunately, for the vdW EoS, an analytical expression of the Widom line in the pressure and temperature plane can be derived from Eq.(46) [21]. By the way, there is another characteristic line defined by the maxima of the heat capacity in a constant temperature process [21]. For the vdW EoS, the later case leads to a linear dependence of pressure on temperature, which is against the findings in real substances, even simple fluids [23] and therefore it is not discussed here. All the details from Eq.(46) are provided in Appendix C. The analytical expression allows us to calculate the derivatives of the pressure and temperature w.r.t. density alone the Widom line. Meanwhile, the derivatives can be obtained for the M-line from Eq.(8), Eq.(9) and Eq.(12). At the critical point we have (Appendix C):

$$\left.\frac{dT_r}{d\rho_{rM}}\right|_C = \left.\frac{dT_r}{d\rho_{rW}}\right|_C = \frac{5}{4} \quad (47)$$

$$\left.\frac{dP_r}{dv_{rM}}\right|_C = \left.\frac{dP_r}{dv_{rW}}\right|_C = -5 \quad (48)$$

$$\left.\frac{dP_r}{dT_{rM}}\right|_C = \left.\frac{dP_r}{dT_{rW}}\right|_C = 4 \quad (49)$$

where the subscript "C" refers to the critical point, "W" to the Widom line, "M", the M-line. Eq.(47)-Eq.(49) are remarkable: they provide rigorous relations between the Widom line and the M-line (the mean-field interface) at the critical point. These equations tell that the continuation of the Widom line with the M-line is at $C^1$ level. For example, in the temperature-pressure plane: $P_{rM}(T_c) = P_{rW}(T_c)$, $\left.\frac{dP_r}{dT_{rM}}\right|_C = \left.\frac{dP_r}{dT_{rW}}\right|_C$. We can also show that the 2nd derivatives are not equal (see Figure C1).

**Results and discussions**

All calculation results are presented by Figure 1 – Figure 7, where the position variable is denoted as $z' = z - z_0$ for brevity. With this notation, a function appeared in previous sections $f(z)$ can be read as $f(z')$ and the derivatives are the same since $dz = dz'$. Figure 1a depicts definitions of the classic (the algebraic mean) Gibbs dividing interface, Eq.(18), and the mean-field interface, Eq.(33) and Eq.(13). In this work, the thickness of the interface will not be addressed and the interface area (between the dashed blue lines) is illustrative only. The mean-field dividing interface correctly reflects the fact that the boundary layer at the vapor side is thinner than that at the liquid side. In contrast, the algebraic mean interface equally divides the interface area, which contradicts physical expectations.

Figure 1b presents the 1st and 2nd derivatives obtained from the mean-field density profile, Eq.(34) and the classic density profile, Eq.(31) at $T^* = 0.15$. This is an important comparison. The position dependences obtained from the two models are different, resulting from the different definitions of the dividing interface. In particular, from Eq.(15) and Eq.(26) we see that there is a relation between the pressure difference and the first derivative of local density:

$$p_N(z) - p_T(z) = m\left(\frac{d\rho}{dz}\right)^2 = 2(\Omega + P^s) \qquad (50)$$

Eq.(5) tells that the inaccuracy in the derivative will cause the inaccuracy in predicting the pressure difference. The new model, Eq.(34), is not only physically favorable, but is more accurate than the classic model, Eq.(31), as shown by Figure 2. Eq.(50) and Figure 1b explain why the classic model, Eq.(31), fails at predicting the pressure difference [15].

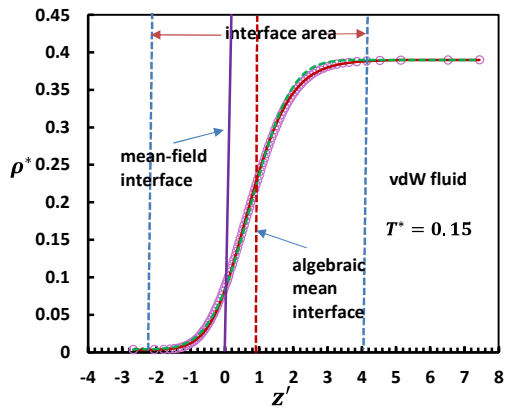

**Figure 1a**. Density profiles at $T^* = 0.15$, where $z' = z - z_0$. The interfacial region is defined illustratively between two dashed blue lines. The red solid curve is from Eq.(34), the dashed green line from Eq.(31) and the circles are from the most recent simulation data [29].

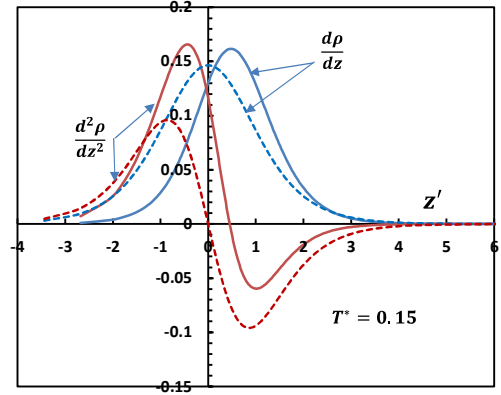

**Figure 1b**. Derivatives from two density profile models: solid lines are from Eq.(34) and dashed lines from Eq.(31).

**Figure 1**. Density profiles and their derivatives at $T^* = 0.15$.

Figure 2 shows detailed correlation results for the density profiles over the entire temperature range, $0.15 \leq T^* \leq 0.25$, by the classic model, Eq.(31) (dashed lines), and by the new model, Eq.(34) (solid lines). The parameters are given by Eq.(35) for the new model, and by Eq.(36) for the classic model, Eq.(31). The results show that the new model, Eq.(35), works excellently over the entire range from vapor phase to liquid phase, while the classic model, Eq.(31), is less accurate, in particular on the liquid side (Figure 2b).

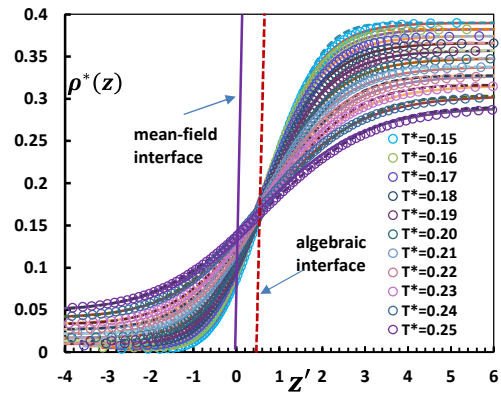

**Figure 2a**. Plots of density profiles in the entire range.

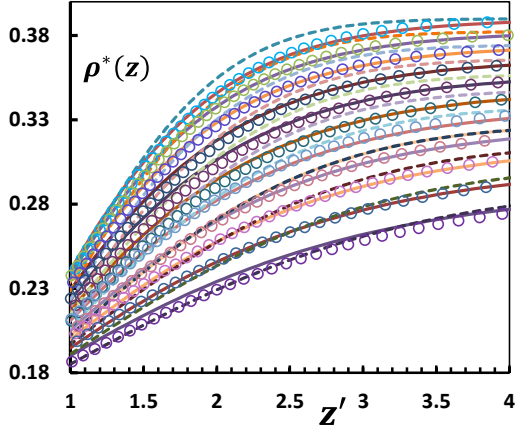

**Figure 2b**. Plots of density profile correlations: enlarged portion on the liquid side.

**Figure 2**. Plots of density profile correlations. Solid lines: Eq.(34); dashed lines: Eq.(31); points: Ref. [29].

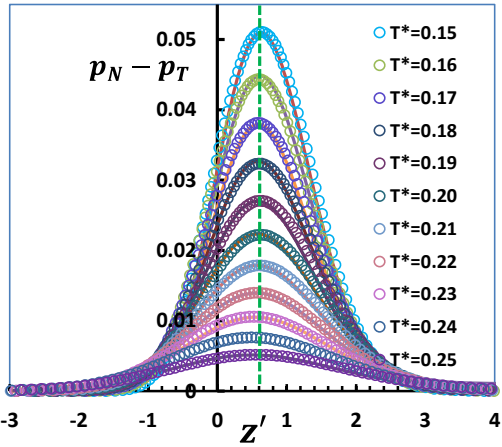

**Figure 3**. Plots of predictions of the pressure difference, $[p_N(z) - p_T(z)]$ by Eq.(50) and Eq.(34) (lines), compared with the simulation data [29].

Figure 3 depicts the prediction results for the pressure difference, $p_N(z) - p_T(z)$, by Eq.(50) combined with Eq.(34). The simulation values and predictions show that $p_N(z) - p_T(z)$ peaks at $z' \approx 0.6$. In summary, the results shown by Figure 2 and Figure 3 demonstrate that the mean-field theory works excellently with the density profile, Eq.(34). By the way, as shown by Eq.(15), a maximum of the pressure difference implies that the change rate (w.r.t position) of the local surface tension peaks at the same position, $z' \approx 0.6$. The highly accurate correlation and prediction guarantee that the applications of Eq.(44) and Eq.(45) will produce reliable results within the framework of the density gradient theory.

Figure 4a depicts the Gibbs free energies across the interface area at $T^* = 0.15$. The figure shows that at $z' \approx 0.6$, there exhibits a maximum with the intrinsic Gibbs free energy, Eq.(44) with Eq.(34), while the classic model, Eq.(31) (with Eq.(44)), fails at predicting the behavior. An important observation is that the Maxwell construction or the equal-area rule holds for the homogeneous contribution where the density $\rho^*(z)$ is used in the vdW EoS (dotted line). It is found that in Eq.(29) the term $\frac{1}{2}m'\left(\frac{d\rho}{dz}\right)^2$ can be neglected without notably impacting the final results. In other words, the impact of density-dependence of the influence parameter is indeed minor.

Figure 4b illustrates the intrinsic Gibbs free energy at different temperatures and it is found that all curves peak at $z' \approx 0.6$. By comparing Figure 4b with Figure 3, we see that a maximum of the pressure difference, $p_N(z) - p_T(z)$, corresponds to a maximum of the intrinsic Gibbs free energy. For a stable fluid the surface tension is directly related to the Gibbs free energy macroscopically [1]. In comparison, for the unstable region Figure 3 and Figure 4b show that the intrinsic Gibbs free energy is related to the change rate of the local surface tension [1,7], $d\gamma = -\Gamma dG$.

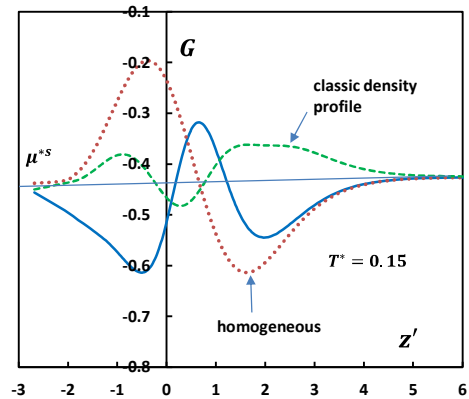

**Figure 4a**. A comparison of various Gibbs free energies at $T^* = 0.15$. The solid blue line is from Eq.(44) with the density profile from Eq.(34), and the dashed green line from classic density model, Eq.(31). The dotted line shows the contribution from the homogeneous mean-field EoS, $\mu^*(\rho(z))$. The horizontal solid line represents the equilibrium value, $\mu^{*S}$.

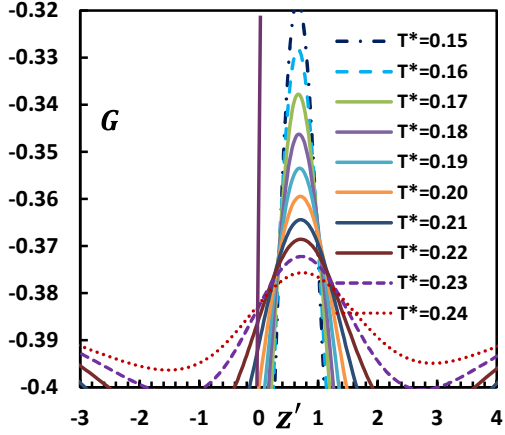

**Figure 4b.** The intrinsic Gibbs free energy calculated by Eq.(44) at different temperatures. The maximum values are located at $z' = 0.5$. The density profile is from Eq.(34).

**Figure 4.** Plots of the Gibbs free energy.

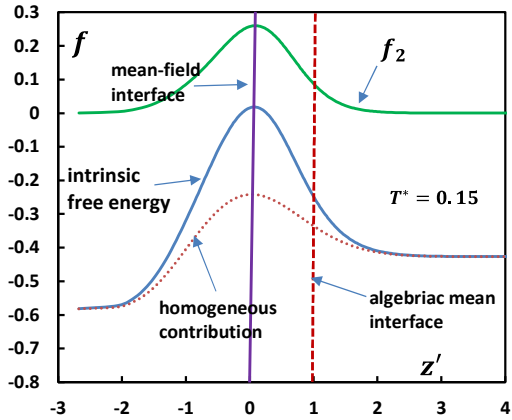

**Figure 5a.** Plots of the Helmholtz free energies with the new density profile at T*=0.15. The intrinsic free energy (solid blue line) is calculated by Eq.(45). The homogeneous contribution is from Eq.(43) with the density $\rho(z)$ by Eq.(34). The green line is the heterogeneous contribution. $f_2 = \tilde{f}_2 T^*$, Eq.(45).

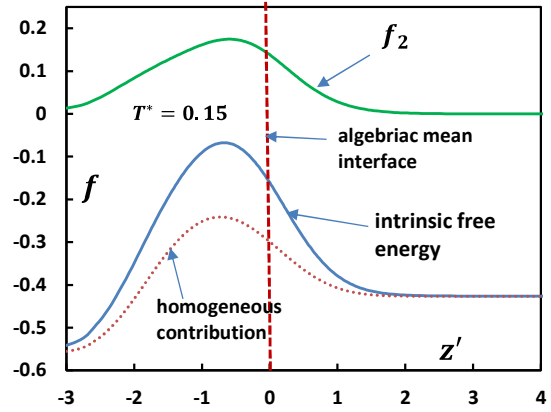

**Figure 5b.** Plots of the Helmholtz free energies with the classic density profile at T*=0.15. The intrinsic free energy (solid blue line) is calculated by Eq.(45). The homogeneous contribution is from Eq.(43) with the density $\rho(z)$ by Eq.(31). The green line is the heterogeneous contribution.

**Figure 5** The Helmholtz free energy at $T^* = 0.15$ from the new density profile, Eq.(34), Fig. 5a, and the classic density profile, Eq.(31), Fig.5b, respectively.

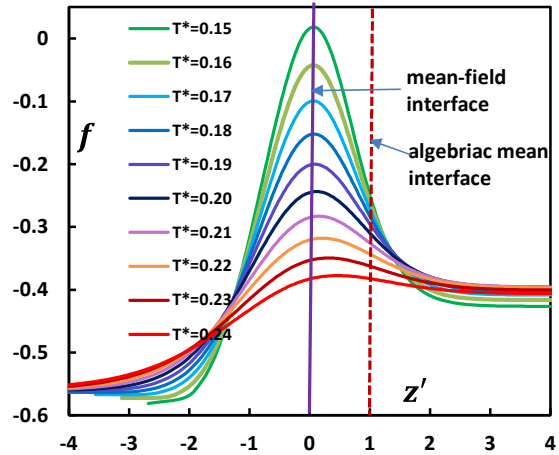

**Figure 6.** The intrinsic free energy, Eq.(45), at various temperature with the new density profile, Eq.(34).

Figure 5 depicts the Helmholtz free energies at $T^* = 0.15$. From Figure 5a, we see that with the new density profile, Eq.(34), the intrinsic free energy peaks at the mean-field dividing interface, namely a state at the interface is the most unstable one. For a comparison, Figure 5b illustrates the free energies when the classic density profile, Eq.(31), is used. An immediate observation is that in the later case the local maximum location of the free energy does not coincide with the dividing interface.

Figure 6 depicts the intrinsic free energy at various temperature. The mean-field interface ($z = z_0$) turns out to the most unstable line. This is an intriguing finding that physically justifies the definition of the interface. As $z > z_0$, liquid drops start to form and eventually entirely liquefies; as $z < z_0$, gas bubbles start to generate and the system totally becomes gas phase. Therefore, the mean-field interface divides the coexistence region into gas-like and liquid-like sub-regions and it is physically coherent with the Widom line in the supercritical region. The mean-field interface also suggest a new perspective for studying the phase decomposition behavior.

In summary, with the mean-field dividing interface the density gradient theory predicts the following interfacial behaviors: (1) maximum intrinsic Helmholtz free energy at the interface; (2) maximum Gibbs free energy in the liquid side, in accordance with the maximum pressure difference, $p_N(z) - p_T(z)$. These features are not found by using the classic density profile model, Eq.(31).

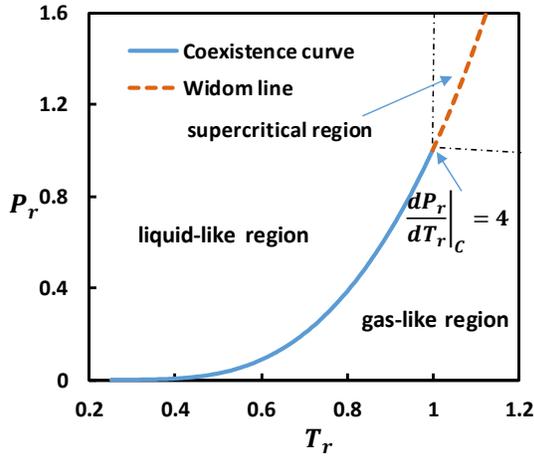

**Figure 7a.** Phase diagram in the ($P_r \sim T_r$) plane.

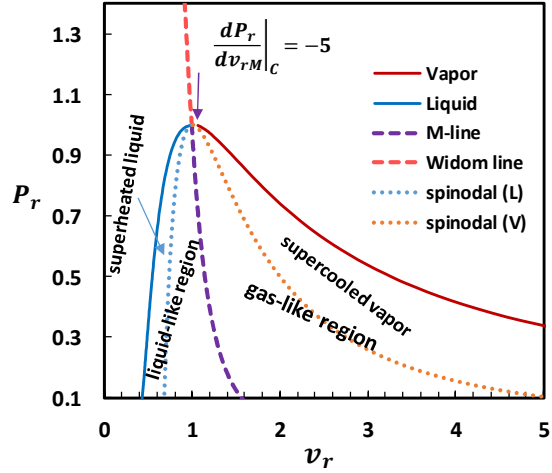

**Figure 7b**. Phase diagram in the ($P_r \sim v_r$) plane

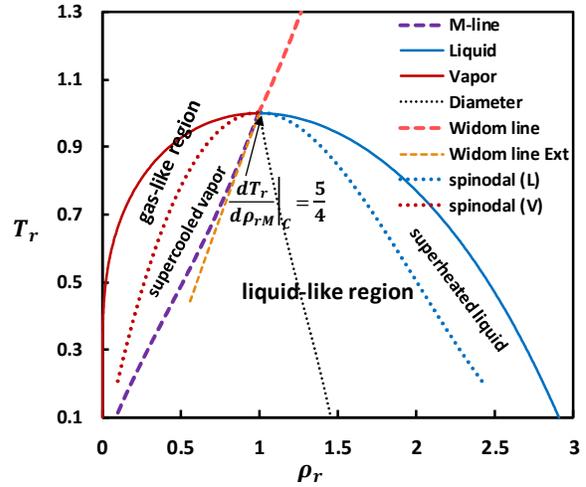

**Figure 7c**. Phase diagram in the ($T_r \sim \rho_r$) plane.

**Figure 7**. Complete phase diagrams in all temperature-pressure-volume planes. The solid lines are from the VLE solutions, Eq.(4)-Eq.(6). The M-line (the mean-field interface) is from Eq.(13), and Widom line from Eq.(C1) and Eq.(C4). Spinodal curves are calculated from $(\partial P/\partial v)_T = 0$.

Figure 7a depicts the phase diagram in the pressure-temperature plan. It is shown that the continuation of the Widom line with the M-line (the mean-field interface) is at $L^1$ level: $P(T_c^+) = P(T_c^-)$, $(dP/dT)_{T_c^+} = (dP/dT)_{T_c^-}$. It can also be shown that the second-order derivatives are not equal (Figure C1).

Figure 7b plots the phase diagram in the pressure-volume plane. The Widom line is calculated in two steps: (a) calculating the temperature at a given pressure using the analytical equation, Eq.(C1); (b) solving the volume with the vdW EoS, Eq.(C4). In a single phase including the supercritical region, the vdW EoS has only one root. It is seen that in the pressure-volume plane the continuation of the Widom line with the M-line is also at $C^1$ level, Eq.(48).

Finally, Figure 7c illustrates the phase diagram in the temperature-density plane. The thin dotted line is the diameter, $d_\sigma = (\rho_{rL} + \rho_{rv})/2$. In calculation of the derivatives, the relation $dT/d\rho = (d\rho/dT)^{-1}$ is applied since we are dealing with the full derivatives. Again, in the temperature-density (volume) plane, the continuation of the Widom –line with the M-line is at $C^1$ level, Eq.(47). In summary, Figure 7 suggests that the continuation of the M-line with the Widom line is at $C^1$ level in all the planes and the liquid-like and gas-like features in the supercritical region are inherited from the coexistence region.

## Conclusions

This paper reports a thorough investigation of the relationship between and the vapor-liquid coexistence region and the supercritical for the van der Waals fluid. The vapor-liquid interface system is explored with the density gradient theory incorporated with the mean-field theory. We show that solving a VLE problem with a cubic EoS is reduced to solving a one-unknown non-linear equation and some simple relations have been found between the saturated volumes, pressure and temperature. A highly accurate expression for the density profile is proposed in which the mean-field Maxwell crossover, or the M-line, is adopted as the dividing interface for the interfacial region. The new expression separates the vapor side of the interface region from the liquid side, which correctly reflects the fact that the decay length of the vapor layer is less than that of the liquid layer [9,15]. The high accuracies of the density-profile correlation and the predictions of the pressure difference, $p_N(z) - p_T(z)$ guarantee that the interfacial properties calculated are reliable.

The results show that at the mean-field interface the intrinsic Helmholtz free energy exhibits a maximum value, namely, the states at the interface are the most unstable ones. Therefore using the mean-field crossover as the dividing interface is physically favorable and coherent with the Widom line in the supercritical region. This finding also resolves the long-standing controversy on the existence of a unique separatrix to dividing the supercritical region into gas-like and liquid-like sub-regions. Indeed, the Widom line defined by Eq.(46) pays such a role. In other words, the two-fluid behavior of the supercritical fluid is inherited from the coexistence. By the way, the intrinsic Gibbs free energy peaks at $z - z_0 \approx 0.6$, corresponding to the same behavior of the pressure difference, $p_N(z) - p_T(z)$.

This work also reveals a new perspective of the mean-field cubic EoS subjected to the Maxwell construction for the VLE calculations. All the information (saturated properties and the dividing interface) required for composing the phase diagrams in the entire phase space is embedded in the theory. Three solutions have their respective roles or significances: two as the saturated volumes of vapor and liquid phases, which are related to the first-order transition and the third one as the dividing interface, which is related to the second-order transition as shown by Eq.(12). As a result, the consistent phase diagrams in all three panes are completed.

As mentioned, for the vdW fluid the analytical expressions of the Widom line, Eq.(C1) and Eq.(C4), make it possible for us to derive rigorous equations, Eq.(47)-Eq.(49), that facilitate our discussions. For most EoS available, the Widom line has to be evaluated with the numerical solution to Eq.(46), namely from the maxima of the isobaric heat capacity in a constant pressure process. The author has employed this method to the Lennard-Jones system and the results are presented elsewhere [32].

## Acknowledgments

The author is grateful to Dr. Mejía for helpful discussions and kindly providing unpublished data of the density profile and the pressure differences, $p_N(z) - p_T(z)$ for the vdW fluid in the temperature range $0.15 \leq T^* \leq 0.25$ ($0.5 \leq T_r \leq 0.84$).

# Appendices

### A. Properties of the van der Waals Fluid

With the scaled units, the vdW EoS is given by [30]:

$$P^* = \frac{T^*}{v^* - b} - \frac{a}{v^{*2}} \quad (A1)$$

The parameters are related to the critical constants:

$$b = 1/(3\rho_C^*), \quad a = 27 P_c^* b^2 \quad (A2)$$

The critical constants used in this work are listed in Table A1 [29].

**Table A1 the critical constants**

| $T_c^*$ | $P_c^*$ | $\rho_c^*$ |
|---|---|---|
| 0.29631 | 0.017684 | 0.15915 |

The reduced Helmholtz free energy is given by [30]:

$$f_r = -\frac{8T_r}{3}\left[\frac{3}{2}\ln T_r + \ln(3v_r - 1) + 1\right] - \frac{3}{v_r} \quad (A3)$$

where $T_r = T^*/T_c^*$. The full derivative of the equilibrium pressure, Eq.(6):

$$(v_{rv} - v_{rL})\frac{dP_r}{dT_r} = \frac{8}{3}\ln\left(\frac{3v_{rv}-1}{3v_{rL}-1}\right) -$$
$$\left[\frac{1}{v_{rv}v_{rL}}\left(3 - \frac{1}{v_{rL}} - \frac{1}{v_{rv}}\right) - \frac{8T_r}{3v_{rv}-1} + \frac{3}{v_{rv}^2}\right]\frac{dv_{rv}}{dT_r}$$
$$+\left[\frac{1}{v_{rv}v_{rL}}\left(3 - \frac{1}{v_{rL}} - \frac{1}{v_{rv}}\right) - \frac{8T_r}{3v_{rL}-1} + \frac{3}{v_{rL}^2}\right]\frac{dv_{rL}}{dT_r} \quad (A4)$$

where $\frac{dv_{rL}}{dT_r}$ and $\frac{dv_{rv}}{dT_r}$ can be obtained from Eq.(8), respectively. Based on a parametric solution [24] to the VLE problem, the following serial expansion at the critical point are obtained [30]:

$$\rho_{rv} = 1 - 2t_0^{\frac{1}{2}} + \frac{2t_0}{5} + \frac{13t_0^{\frac{3}{2}}}{25} + \frac{128}{875}t_0^2 + \cdots \quad (A5)$$

$$\rho_{rL} = 1 + 2t_0^{\frac{1}{2}} + \frac{2t_0}{5} - \frac{13t_0^{\frac{3}{2}}}{25} + \frac{128}{875}t_0^2 + \cdots \quad (A6)$$

$$\rho_{rv} + \rho_{rL} = 2 + \frac{4t_0}{5} + \frac{256}{875}t_0^2 + \frac{272}{3125}t_0^3 + \cdots \quad (A7)$$

where $t_0 = 1 - T_r$, $dt_0/dT_r = -1$. From Eq.(A7) and Eq.(9) we have

$$\rho_{rM} = 3 - (\rho_{rv} + \rho_{rL}) = 1 - \frac{4t_0}{5} - \frac{256}{875}t_0^2$$
$$- \frac{272}{3125}t_0^3 + \cdots \quad (A8)$$

Eq.(A8) contains exact leading derivatives. As the three leading terms are used, a highly accurate correlation can be obtained, by fitting the exact (numerical) solutions to Eq.(7), which is Eq.(12). From Eq.(A8) we have:

$$\left.\frac{dv_{rM}}{dT_r}\right|_C = -\frac{4}{5}; \quad \left.\frac{d^2v_{rM}}{dT_r^2}\right|_C = 1.86514 \quad (A9)$$

Hence:

$$\left.\frac{d\rho_{rM}}{dT_r}\right|_C = \frac{4}{5}, \quad \left.\frac{dv_{rM}}{dT_r}\right|_C = -\frac{4}{5} \quad (A10)$$

Since here we are discussing a full derivative, we have

$$\left.\frac{dT_r}{d\rho_{rM}}\right|_C = \frac{5}{4}, \quad \frac{dT_r}{dv_{rM}} = -\frac{5}{4} \quad (A11)$$

Finally:

$$\frac{dP_r}{dv_{rM}} = \frac{8}{3v_{rM}-1}\frac{dT_r}{dv_{rM}} + \left[\frac{6}{v_{rM}^3} - \frac{24T_r}{(3v_{rM}-1)^2}\right]\frac{dv_{rM}}{dT_r} \quad (A12)$$

and:

$$\left.\frac{dP_r}{dv_{rM}}\right|_C = -5 \quad (A13)$$

For the low temperature range, $T_r < 0.3$, $v_{rv} \gg v_{rL}$, from Eq.(7) and Eq.(8), we can derive the following highly accurate analytical solutions:

$$v_{rL} = \frac{9}{16T_r}\left[1 - \left(1 - \frac{32}{27}T_r\right)^{\frac{1}{2}}\right] \quad (A14)$$

$$v_{rv} = \frac{1}{3}(3v_{rL} - 1)exp\left(1 + \frac{3v_{rL}}{3v_{rL}-1}\right) \quad (A15)$$

Eq.(A14), Eq.(A15) and Eq.(9) give $v_{rM}$. Therefore we have explicit expressions for all the saturated properties in the entire temperature range: as $T_r < 0.3$ using Eq.(14) and Eq.(15) to calculate $v_{rL}$ and $v_{rv}$, respectively; as $T_r \geq 0.3$ using Eq.(13) to calculate $v_{rM}$, Eq.(8) to calculate $v_{rL}$ and $v_{rv}$. In all cases, Eq.(6) is used to calculate the pressure.

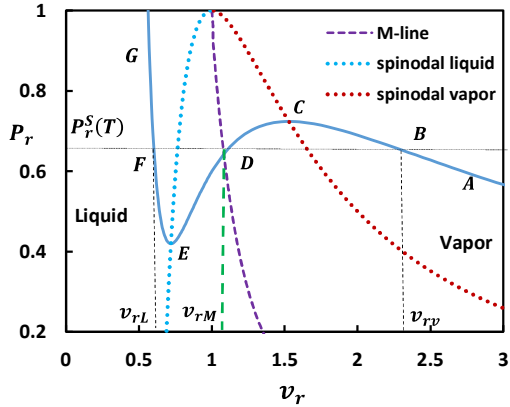

**Figure A1**. The Maxwell construction and the M-line. $v_r = v^*/v_c^*$ etc.

Figure A1 illustrates the Maxwell construction and all related properties. $P_r^S(T_r)$ (horizontal dotted line) is the equilibrium pressure at a given temperature. The cubic curve is produced by the vdW EoS at $T_r = 0.9$. A cooling process starts from point *A* in the vapor region, and ends at point *G* in the liquid region. *BC* represents the supercooled vapor phase, *FE*, the superheated liquid phase, and *FB*, the coexistence phase. The intermediate volume, $v_{rM}$, is obtained from the equal-area rule, area FED = area DCB: $\int_{v_{rL}}^{v_{rM}}(P_r^S - P_r)dv_r = \int_{v_{rM}}^{v_{rv}}(P_r - P_r^S)dv_r$.

Figure A2 depicts the process of determining the equilibrium (saturated) pressure, $P^{*S}$, and the chemical potential, $\mu^{*S}$, at a given temperature. These two values are required in predicting the grand potential difference, $\Delta\Omega$, Eq.(23). The equilibrium chemical potential has been correlated with a quadratic function in the temperature range considered (in terms of reduced temperature):

$$\mu^{*S} = -0.57047T_r^2 + 0.89309T_r - 0.73092 \quad (A16)$$

Or $\mu^{*S} = -6.5008T^{*2} + 3.01532T^* - 0.73103$.

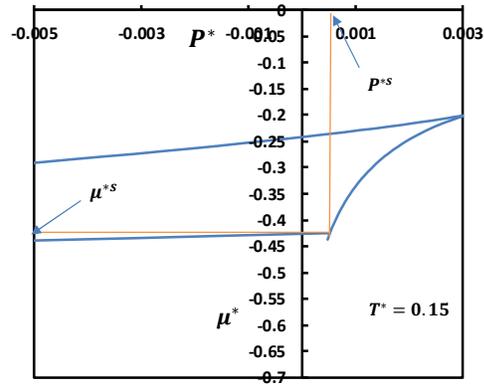

**Figure A2**. The determination of the equilibrium pressure and chemical potential in the pressure-chemical potential plane at a given temperature.

### B. A new density profile model

In this section, we derive a new density profile expression. We start with the governing differential equation for the density profile by taking the vapor phase as the bulk fluid [9]:

$$\alpha_2\left(\frac{d\rho_v(z)}{dz}\right)^2 + \alpha_0[\rho_v^2(z) - \rho_v^2] + 2(1 - \alpha_0\rho_v)[\rho_v(z) - \rho_v] = 2\rho_v(z)ln\frac{\rho_v(z)}{\rho_v} \quad (B1)$$

Eq.(B1) is obtained by truncating an Taylor expansion at the 2nd order while the equation is exact up to the 3rd order (odd derivative terms being zero) by omitting the

4th and higher orders. The coefficients in eq.(B1) are defined as [7,9]:

$$\alpha_0 = \frac{4\pi\beta}{3}\int_0^\infty r^3 g(\rho_b; r)\frac{du(r)}{dr}dr \quad (B2)$$

$$\alpha_2 = \frac{2\pi\beta}{15}\int_0^\infty r^5 g(\rho_b; r)\frac{du(r)}{dr}dr \quad (B3)$$

where $\alpha_0 > 0$, $\alpha_2 > 0$. This feature is of critical importance for solving Eq.(B1). We define a new variable:

$$\bar{\rho} = \rho_v(z) - \rho_v > 0 \quad (B4)$$

By considering the fact that the above differential equation, Eq.(B1), is exact up to the 3rd order, the expansion of the logarithm term ought to be written as:

$$ln\frac{\rho_v(z)}{\rho_v} = ln\left(\frac{\bar{\rho}}{\rho_v}+1\right) \approx \frac{\bar{\rho}}{\rho_v} - \frac{1}{2}\frac{\bar{\rho}^2}{\rho_v^2} + \frac{1}{3}\frac{\bar{\rho}^3}{\rho_v^3} \quad (B5)$$

Here we are expanding the log term as a scalar hence all odd terms should be kept. This is where the current work is different from Ref. [9], in which the third order is omitted and hence the accuracy does not match that of Eq.(B1). By inserting Eq.(B5) into Eq.(B1), after some rearrangements, we obtain:

$$\left(\frac{d\bar{\rho}}{dz}\right)^2 = b\bar{\rho}^2 - a\bar{\rho}^3 \quad (B6)$$

where

$$a = \frac{1}{3\alpha_2}\frac{1}{\rho_v^2} > 0, \quad b = \frac{1}{\alpha_2}\left(\frac{1}{\rho_v} - \alpha_0\right) > 0 \quad (B7)$$

The last arguments, Eq.(B7), need some justifications since they are critical for obtaining the final solution. From the definitions, Eq.(B2), Eq.(B3), we can estimate the values of $\alpha_0$ and $\alpha_2$ when a radial distribution function (RDF) is known. However, there is no exact expression for the DRF of the vdW fluid. Here we assume that the Lennard-Jones (LJ) can be considered to be a close approximation [30], for which the RDF expressions are available [33]. By using this analytical expression [33], some computations have been carried out for the LJ fluid and the results show that Eq.(B7) holds. The calculations also show that "$b$" and $a\bar{\rho}$ are in the same order of magnitude and thus ignoring the term $a\bar{\rho}^3$ in Eq.(B5) (as did in Ref.[9]) is not acceptable. For integrating Eq.(B6), a new variable is introduced: $u = \sqrt{b - a\bar{\rho}}$, and by noticing that $a\bar{\rho} = b - u^2 > 0$, Eq.(A6) can be written as:

$$-\frac{1}{2}\frac{d\bar{\rho}}{\bar{\rho}\sqrt{b-a\bar{\rho}}} = \frac{du}{b-u^2} = -\frac{1}{2}dz \quad (B8)$$

On integration of Eq.(B8) [34], we have:

$$\int\frac{du}{b-u^2} = \frac{1}{\sqrt{b}}atanh\left(\frac{u}{\sqrt{b}}\right) + C = \frac{1}{2}(z-z_0) \quad (B9)$$

In the current model, we define $z_0$ in such a manner that the region $-\infty < (z-z_0) < 0$ is for the vapor side, and $0 < (z-z_0) < \infty$ for the liquid side. Therefore at the vapor side, $z < z_0$, and in Eq.(B9), the sign has been changed so that the same shifted variable $z - z_0$ can be applied to both liquid and vapor sides at the same time. By using the properties of the hyperbolic functions [34], Eq.(B9) can be rewritten as:

$$u = \frac{\tanh(-C\sqrt{b}) + thanh\left[\frac{\sqrt{b}}{2}(z-z_0)\right]}{1+\tanh(-C\sqrt{b})thanh\left[\frac{\sqrt{b}}{2}(z-z_0)\right]} \quad (B10)$$

The conditions for obtaining Eq.(B10) are $b > 0$ and $b - u^2 > 0$ [34] and the definitions of $b$ and $u$ guarantee that both conditions are met as discussed above. Since $|\tanh(x)| < 1$, we can neglect all the terms with $\tanh^2(x)$ and finally after some straightforward algebra we have:

$$\rho(z) = A + \frac{B\tanh\left(\frac{z-z_0}{D}\right)}{1+C\tanh\left(\frac{z-z_0}{D}\right)} \quad (B11)$$

where $A$, $B$ and $C$ are (bulk) density-dependent, and $D = 2/\sqrt{b}$, which is a quantity related to the thickness of the interfacial region. Eq.(B11) is derived for the vapor side of the interface ($\rho(z)$ should be read as $\rho_v(z)$). For obtaining a single density profile equation, we empirically "merge" the vapor side and liquid side with the following form:

$$\rho(z) = A + \frac{B\tanh\left(\frac{z-z_0}{D_v}\right)}{1+C\tanh\left(\frac{z-z_0}{D_L}\right)} \quad (B12)$$

where $D_v$ is related to the thickness of the vapor layer, and $D_L$ to that of the liquid layer. Hence, $D_v + D_L$ is related to the total thickness of the interface area, corresponding to the parameter $D$ in the classic model, Eq.(31). The three parameters, A, B and C can be determined by the conditions discussed in the main text. From Eq.(B12), the derivatives can be easily obtained:

$$\frac{d\rho(z)}{dz} = \frac{B \operatorname{sech}^2 \frac{z-z_0}{D_v}}{1 + C \tanh \frac{z-z_0}{D_L}} \left[ \frac{1}{D_v} - \frac{1}{D_L} \frac{C \tanh \frac{z-z_0}{D_v}}{1 + C \tanh \frac{z-z_0}{D_L}} \right] \quad (B13)$$

$$\frac{d^2\rho(z)}{dz^2} =$$
$$-2 \frac{d\rho(z)}{dz} \left( \frac{1}{D_v} \tanh \frac{z-z_0}{D_v} + \frac{1}{D_L} \frac{C \operatorname{sech}^2 \frac{z-z_0}{D_v}}{1 + C \tanh \frac{z-z_0}{D_L}} \right) \quad (B14)$$

where:

$$B = 2(\rho_L - \rho_M)\left(\frac{\rho_M - \rho_v}{\rho_L - \rho_v}\right) \quad (B15)$$

$$C = \frac{2\rho_M - \rho_v - \rho_L}{\rho_L - \rho_v} \quad (B16)$$

Eq.(A12)-(A14) are used for calculations of density gradients discussed in the main text.

**C. The Widom line in the $(P\sim v)$ and $(T\sim v)$ planes**

The analytical expression of the Widom line in $(T_r\sim P_r)$ plane has been derived by Lamorgese et al.[21]:

$$T_r = \frac{1}{16}\left(-1 + \frac{P_r}{W} + \frac{W}{P_r}\right)\left[P_r + 108\left(1 + \frac{P_r}{W} + \frac{W}{P_r}\right)^{-2}\right] \quad (C1)$$

where

$$W = \left[6P_r^2\sqrt{3(27 + P_r)} + P_r^2(54 + P_r)\right]^{\frac{1}{3}} \quad (C2)$$

This will provide the $P_r\sim T_r$ curve in the supercritical region. For the volume calculations, the vdW EoS, Eq.(1), is written as:

$$3P_r v_r^3 - (P_r + 8T_r)v_r^2 + 9v_r - 3 = 0 \quad (C3)$$

This function has only one root as $T_r > 1$. Defining $\Delta_0 = (P_r + 8T_r)^2 - 81P_r$, $\Delta_1 = -2(P_r + 8T_r)^3 + 243P_r(P_r + 8T_r) - 729P_r^3$, and $C = \left[\frac{1}{2}\left(\Delta_1 + \sqrt{\Delta_1^2 - 4\Delta_0^3}\right)\right]^{\frac{1}{3}}$, we have

$$v_{rW} = -\frac{1}{9P_r}\left[-(P_r + 8T_r) + C + \frac{\Delta_0}{C}\right] \quad (C4)$$

where the subscript "W" refers to the Widom line. Eq.(C1) and (C4) provide the Widom line in the $(P_r\sim v_r)$ and $(T_r\sim v_r)$ planes, respectively. For the derivatives, we define:

$$A = \frac{1}{W} - \frac{P_r}{W^2}\frac{dW}{dP_r} - \frac{W}{P_r^2} + \frac{1}{P_r}\frac{dW}{dP_r} \quad (C5)$$

$$B = 1 + \frac{P_r}{W} + \frac{W}{P_r}, \quad C = \sqrt{3(27 + P_r)} \quad (C6)$$

Then we have

$$\frac{dW}{dP_r} = \frac{1}{3}[6P_r^2 C + P_r^2(54 + P_r)]^{-\frac{2}{3}}$$
$$\left[12P_r C + \frac{9P_r^2}{C} + 2P_r(54 + P_r) + P_r^2\right] \quad (C7)$$

and finally:

$$\frac{dT_r}{dP_r} = \frac{A}{16}\left(P_r + \frac{108}{B^2}\right) + \frac{1}{16}(B - 2)\left(1 - \frac{216A}{B^3}\right) \quad (C8)$$

At the critical point:

$$\left(\frac{dP_r}{dT_r}\right)_{WC} = 4 \quad (C9)$$

And

$$\frac{dv_{rW}}{dT_r} = \frac{1}{9P_r^2}\frac{dP_r}{dT_r}\left[-(P_r + 8T_r) + C + \frac{\Delta_0}{C}\right]$$
$$-\frac{1}{9P_r}\left[-\left(\frac{dP_r}{dT_r} + 8\right) + \left(1 - \frac{\Delta_0}{C^2}\right)\frac{dC}{dT_r} + \frac{1}{C}\frac{d\Delta_0}{dT_r}\right] \quad (C10)$$

$$\left.\frac{dv_{rW}}{dT_r}\right|_C = -\frac{4}{5} \quad (C12)$$

$$\frac{dv_{rW}}{dP_r} = \frac{1}{9P_r^2}\left[-(P_r + 8T_r) + C + \frac{\Delta_0}{C}\right]$$
$$-\frac{1}{9P_r}\left[-\left(8\frac{dT_r}{dP_r}\right) + \left(1 - \frac{\Delta_0}{C^2}\right)\frac{dC}{dP_r} + \frac{1}{C}\frac{d\Delta_0}{dP_r}\right] \quad (C13)$$

$$\left.\frac{dv_{rW}}{dP_r}\right|_C = -\frac{1}{5} \quad (C14)$$

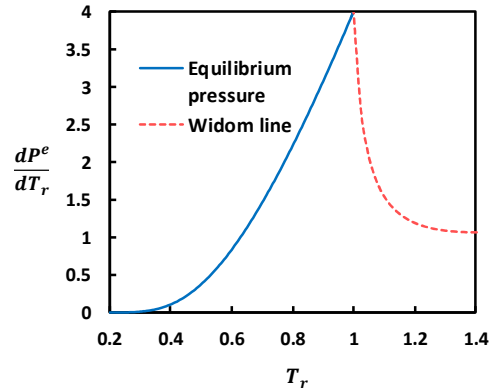

**Figure C1**. Full derivative of the equilibrium pressure with respect to temperature. The solid line for the equilibrium pressure is calculated by Eq.(A4). The derivative of the Widom line is calculated with Eq.(C8).

Figure C1 shows the derivatives of the pressure in the temperature-pressure plane. It is seen that at the critical point the first derivative of the equilibrium pressure equals to that of the Widom line while the second derivatives of the pressure are NOT equal. The same behaviors are found in the temperature-volume and pressure-volume planes. Therefore, the continuation of the M-line and the Widom line is at $C^1$ level in all the planes.